\documentstyle[11pt,paspconf]{article}

\begin{document}

\title{
Submillimeter-selected galaxies 
}

\author{A. W. Blain}
\affil{Cavendish Laboratory, Madingley Road, Cambridge, CB3 0HE, UK
}

\author{Ian Smail} 
\affil{Department of Physics, University of Durham, South Road, 
Durham, DH1 3LE, UK} 

\author{R. J. Ivison} 
\affil{Department of Physics \& Astronomy, University College London, 
Gower Street, London, WC1E 6BT, UK}

\author{J.-P. Kneib} 
\affil{Observatoire Midi-Pyr\'en\'ees, 14 Avenue E. Belin, 
31400 Toulouse, France} 




\begin{abstract}
The first generation of submillimeter(submm)-wave surveys are being 
carried out using the 450/850-$\mu$m SCUBA camera at the JCMT on 
Mauna Kea. These surveys are potentially sensitive to galaxies at very 
high redshift, and the galaxies that have been detected so far appear to 
contribute the greater fraction of the mm/submm-wave background 
radiation intensity measured by {\it COBE}. In order 
to understand this new population of galaxies, individual examples must 
be studied in detail across many wavebands; in particular their 
redshifts must be determined. We discuss the potential selection 
effects at work in submm-wave surveys and describe the spectral energy 
distributions (SEDs) of galaxies selected or luminous in the submm 
waveband. We also describe the general procedure for, and 
emphasize the difficulty of, identifying optical counterparts to
submm-selected galaxies. Finally, we summarize what is known about 
the redshifts of these galaxies
and the source of their luminosity.
\end{abstract}


\keywords{galaxies: formation, galaxies: evolution, cosmology: observations, 
gravitational lensing, galaxies: general}


\section{Introduction}

The first submm-wave surveys, designed to detect the redshifted 
restframe far-infrared(IR) radiation from warm dust in distant 
galaxies (Blain \& Longair 1993), have recently been carried out 
(Smail, Ivison \& Blain 1997; Barger et al.\ 1998, 1999b; Hughes et al.\ 1998; 
Blain et al.\ 1999b, 2000; Eales et al.\ 1999). These 
surveys are uniquely sensitive to high-redshift galaxies, and 
have been possible since the sensitive SCUBA camera 
(Holland et al.\ 1999) was commissioned at the JCMT in 1997. SCUBA 
is an imaging and photometric instrument, optimized to 
observe a 5-arcmin$^2$ field simultaneously in the 450- and 850-$\mu$m 
atmospheric transmission windows. The images produced 
have a resolution of about 15\,arcsec at 850\,$\mu$m and 7\,arcsec at 
450\,$\mu$m. 

The intensity of mm, submm and far-IR background radiation has been 
measured recently using data from the FIRAS and DIRBE instruments on 
the {\it COBE} satellite (Puget et al.\ 1996; Fixsen et al.\ 1998; 
Hauser et al.\ 1998; Schlegel, 
Finkbeiner \& Davis 1998; Finkbeiner, Davis \& Schlegel 1999). The results 
of the first galaxy surveys in the far-IR waveband, carried out 
using the {\it ISO} satellite, have also been reported recently 
(Kawara et al.\ 1998; Puget et al.\ 1999). 

In this paper we review certain aspects of the existing submm-wave surveys, 
concentrating on the selection technique and the procedures for following 
up the submm-wave detections in other wavebands. We discuss whether 
any selection effects could act to prevent any types of high-redshift 
dusty galaxies being detected efficiently in submm-wave 
surveys, and compare all the well-determined  
SEDs of submm-selected galaxies with those of other dusty galaxies and 
active galactic nuclei (AGNs). 
We discuss the difficulties of identifying the galaxy responsible for the 
detected submm emission, and show an example, the SCUBA-selected 
galaxy SMM\,J09429+4658 (Smail et al.\ 1999a), in which very 
deep multiwaveband images of the fields around the centroid of the 
SCUBA source were required in order to make progress in the identification.
Finally, we discuss our current state of knowledge about the redshift distribution 
of the galaxies detected in SCUBA surveys, which is crucial for distinguishing 
between different models of galaxy evolution (Blain et al.\ 1999a,c), about 
the connections between the high-redshift SCUBA sources and the 
properties of dust in low-redshift galaxies, and about whether the 
SCUBA galaxies are powered by star-formation activity or by AGN accretion. 
Unless otherwise stated we assume an Einstein--de Sitter world model  
with Hubble's constant $H_0=50$\,km\,s$^{-1}$\,Mpc$^{-1}$. 

\section{Submm selection effects: spectral energy distributions (SEDs)} 


The reason why a large fraction of the galaxies detected in submm-wave 
surveys are at high redshifts is the steep long-wavelength slope of 
the dust emission spectrum, which can be described well by a 
Rayleigh--Jeans thermal spectrum, modified by an emissivity 
function $\epsilon_\nu$. An emissivity with the form 
$\epsilon_\nu \propto \nu^\beta$, where $\beta \simeq 1.5$ provides a 
reasonable fit to the data, and so 
the submm-wave emission spectrum of a dusty galaxy 
$f_\nu \propto \nu^\alpha$, where $\alpha \simeq 3.5$. This steep 
spectrum leads to a strong negative $K$-correction, which at 
redshifts $z > 0.5$ can be sufficient to overcome the geometric dimming 
due to the inverse square law, and lead to a flat flux density--redshift 
relation in the submm waveband. The SED is also defined by a dust 
temperature $T_{\rm d}$. Low-redshift spiral galaxies generally have 
dust temperatures of about 15--20\,K, but there is evidence, from both
individual high-redshift dusty galaxies and the background 
radiation spectrum, that hotter temperatures, $T_{\rm d} \simeq 40$\,K,  
are typical of the submm-selected galaxy population (see Section\,4), 
at least as a luminosity-weighted mean temperature. 

The strong submm-wave $K$-correction is illustrated in Fig.\,1. 
We show the flux density--redshift relation predicted for a dusty 
galaxy with a fixed bolometric luminosity in the restframe 
far-IR waveband for a variety of dust temperatures $T_{\rm d}$ and 
dust emissivity indices $\beta$. The SED template used is 
assumed to have a power-law slope $f_\nu \propto \nu^{-1.7}$ on the 
Wien side of the dust emission spectrum in the mid-IR waveband. 
This power-law slope takes account of the contribution made to the 
SED at shorter wavelengths by the populations of hotter dust grains 
in the interstellar medium (ISM) of the galaxy, and provide a good 
representation of very deep 15-$\mu$m counts of galaxies determined 
using the {\it ISO} satellite (Altieri et al.\ 1999), see Blain et al.\ (1999a,c) 
for more information. 

\begin{figure}[t]
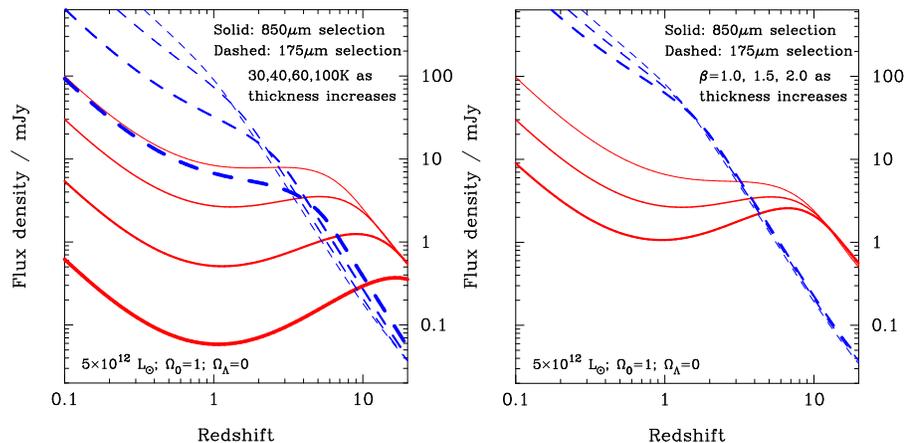

\begin{center}
\plotfiddle{blaina2_1a.ps}{4.4cm}{-90}{38}{38}{-230}{185}
\end{center}
\begin{center}
\plotfiddle{blaina2_1b.ps}{4.3cm}{-90}{38}{38}{-60}{332}
\end{center}
\vskip -5cm
\caption{ 
Flux density--redshift relations for dusty galaxies with fixed 
bolometric luminosities at wavelengths of 850 and 175\,$\mu$m in the 
submm and far-IR wavebands. Left: the emissivity index $\beta = 1.5$ is 
fixed and the dust temperature $T_{\rm d}$ varies. Right: $T_{\rm d}=38$\,K is 
fixed and $\beta$ varies.  
}
\end{figure}

It is clear from the left-hand panel of Fig.\,1 that the dust temperature 
has a significant effect on the detectability of a dusty galaxy with a 
fixed bolometric luminosity, both when observed close to the peak of 
the SED at 175\,$\mu$m in the far-IR waveband and at 850\,$\mu$m in 
the submm waveband. Hotter galaxies produce less submm-wave
flux density per unit bolometric luminosity, and are thus less likely to be 
detected in the submm waveband. This effect is illustrated by a comparison 
of the 850-$\mu$m flux densities, dust temperatures and bolometric 
luminosities of APM\,08279+5255 (Lewis et al.\ 1998) and IRAS F10214+4724 
(Lacy et al.\ 1998). Although the 850-$\mu$m flux density of APM\,08279+5255
is only a factor of 1.5 times greater than that of IRAS F10214+4724, 
because $T_{\rm d} > 100$\,K in APM\,08279+5255 but only 
$\simeq 80$\,K in IRAS F10214+4724, the inferred luminosity of 
APM\,08279+5255 exceeds that of IRAS F10214+4724 by an order 
of magnitude.

It is interesting to note that the dust temperature that best accounts for the 
properties of the faint submm galaxy population (Blain et al.\ 1999a,c; 
Trentham et al.\ 1999) is close to 40\,K, which also seems to be typical of 
the dust in high-redshift quasars and radio galaxies (Benford et al.\ 1999; 
Ivison et al.\ 1998a) and in {\it IRAS} galaxies (Lisenfeld, Isaak \& Hills 1999). 
It is possible therefore that an additional population of high-redshift galaxies 
could exist with hotter dust temperatures, that are underrepresented in 
submm-selected samples. Note that such a population would probably be 
too distant to have been detected using the {\it ISO} satellite at 175\,$\mu$m. 

The effect of varying the emissivity index of the dust spectrum is shown 
in the right-hand panel of Fig.\,1. The value of the index has little effect 
on the detectability of galaxies at a far-IR wavelength of 175-$\mu$m, 
which is close to the restframe peak of the SED; however, the effects are 
more significant at a submm wavelength of 850\,$\mu$m. Low-redshift 
galaxies would be more likely to be detected in a SCUBA survey if the 
emissivity index $\beta$ is low. 

\begin{figure}[t]
\begin{center}
\plotfiddle{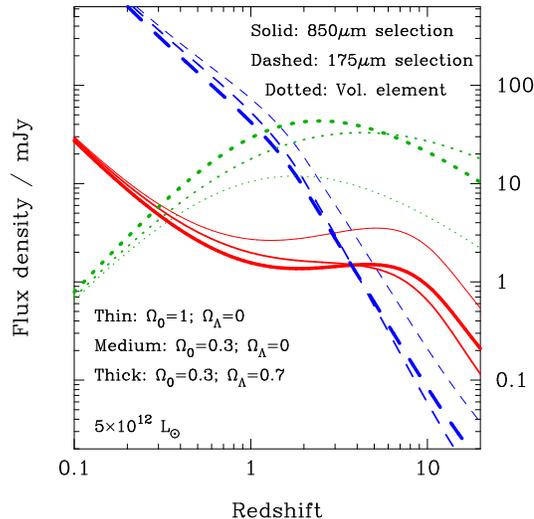}{5.7cm}{-90}{45}{45}{-175}{230}
\end{center}
\caption{The flux density--redshift relation for a dusty 
galaxy with a dust temperature of 38\,K and an emissivity 
index $\beta = 1.5$ in three different cosmological models. 
The relative size of the volume element as a function of 
redshift is also shown in each model, in arbitrary units. 
}
\end{figure} 

\section{Submm selection effects: cosmology} 

Because submm-wave surveys probe the Universe at $z \ge 1$, 
the effects of different world models on the observability of dusty galaxies
could be significant. The effects are illustrated in Fig.\,2, in which the 
flux density--redshift relation for a dusty galaxy with a fixed template 
SED and a fixed bolometric luminosity is compared in three different 
world models. 
The $K$-correction is less dramatic in the two $\Omega_0 < 1$
models as compared with the Einstein--de Sitter model, and so in these
two models, the intrinsic luminosity of a high-redshift dusty galaxy 
detected at a certain 850-$\mu$m flux density will be greater. However, 
the effect of the world model on the interpretation of submm-wave 
source counts is not very significant. This is because the volume element 
is also larger in the two $\Omega_0 < 1$ models, and so, in order to 
avoid overpredicting the measured source counts, the evolution of  
either the space density or luminosity of the population of submm-selected 
galaxies must be 
less dramatic in these models. The effects of an increased luminosity 
per unit flux density and a less dramatic form of evolution will 
largely counteract each other, and so the general conclusions about 
evolution are unchanged.

When accurate counts and redshift distributions are available 
for submm-selected galaxies, and their SEDs are known, it 
will be possible to investigate the effects of different cosmological 
models. However, at present the uncertainties due to the SED are 
greater than those due to the world model.  

\section{The SEDs of submm galaxies with known redshifts} 

In Fig.\,3 we compare the observed restframe SEDs of luminous 
dusty galaxies with known redshifts and two template SEDs derived for 
use in galaxy evolution models. The Guiderdoni et al.\ (1998) model 
was compiled from a large sample of SEDs of {\it IRAS} galaxies, and 
the Blain et al.\ (1999c) model was derived from the counts of {\it IRAS} 
and {\it ISO} galaxies at 60 and 175\,$\mu$m. Observed SEDs for 
several classes of submm-luminous source are shown: low-redshift 
{\it IRAS} galaxies, for which SEDs are available right across the 
submm, far- and mid-IR wavebands; submm-selected galaxies, of 
which four have known redshifts; lensed quasars; and both optical- and 
radio-selected high-redshift AGN. The optical-selected extremely red 
object (ERO) HR10 (Dey et al.\ 1999) is included as a submm-selected 
galaxy, as it would have been identified in the current surveys. 

Both of the model spectra shown in Fig.\,3 appear to provide 
reasonable descriptions of most of the galaxies, and all of 
the submm-selected ones. The exceptions are the sources 
that are known to be strongly lensed by foreground 
galaxies and some of the high-redshift active galaxies, in which the  
dust temperature is considerably higher than 40\,K. 
However, it is reasonable to have a greater mid-IR luminosity 
from these objects, due both to intense heating by an AGN and 
to differential magnification across the nucleus of the source. Both 
of these processes would be expected to increase the 
dust temperature required to fit the observed SED 
(Eisenhardt et al. 1996; Blain 1999a). In general, an SED model 
with $T_{\rm d} \simeq 40$\,K and $\beta \simeq 1.5$ seems to 
describe the observed SEDs of submm-selected dusty galaxies 
reasonably well. 

A lower value of $\beta=1$ was assumed by Eales et al.\ (1999)  
to help explain the significant fraction of galaxies at $z < 1$ 
identified by Lilly et al.\ (1999) in their SCUBA survey of the CFRS fields. 
Very few galaxies are expected to be detected at these low redshifts 
based on models with $\beta=1.5$ (Blain et al.\ 1999c), and in deep 
multiwaveband follow-up observations of fifteen background sources in 
the fields of seven lensing clusters (Smail et al.\ 1997, 1998, 1999a)
none are found at $z<1$. The most appropriate value of $\beta$ will
be clarified by further observations, which will also 
reveal whether some of the low-redshift 
galaxies in the CFRS fields could be mis-identified. 

\subsection{The population of submm-luminous galaxies at low redshifts} 

Faint submm-wave surveys are currently only possible in small fields, 
and so probe very high-redshift pencil beams. It is also interesting to 
investigate the abundance, distribution and properties of dust in 
low-redshift galaxies, which reflects the evolutionary processes at work, 
both star formation and chemical evolution, in the ISM of galaxies 
throughout their history. Dunne et al.\ (1999) have recently 
reported the results of a survey of typical low-redshift {\it IRAS} galaxies, for 
which they obtained 450- and 850-$\mu$m submm-wave SCUBA 
photometry to investigate the dust temperature and spectral index. Note, 
however, that at $z=0$ the SEDs of even normal spiral galaxies, with 
dust temperatures as low as 20\,K, are probed by SCUBA quite a long 
way down the Rayleigh--Jeans slope of the dust SED, and that their 
bolometric luminosity is constrained by the 60-$\mu$m {\it IRAS} data. The 
results will be interesting, but should have no dramatic consequences for the 
interpretation of the high-redshift SCUBA surveys, in which the detected 
galaxies are both observed at significantly shorter restframe wavelengths, 
closer to the peak of their restframe SED, and are much more luminous.  

\begin{figure}[t]
\begin{center}
\plotfiddle{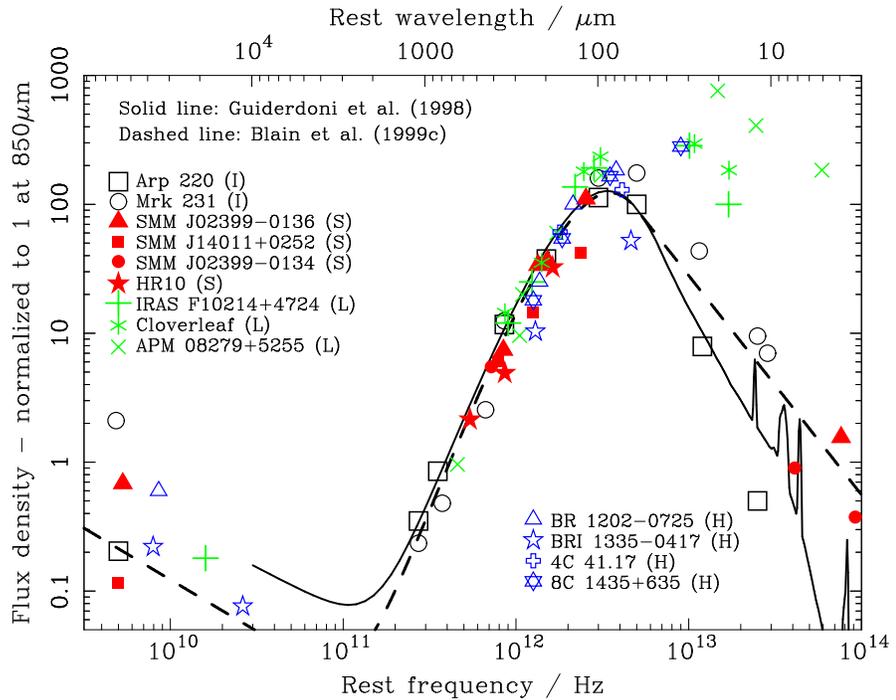}{8.0cm}{-90}{51}{51}{-195}{270}
\end{center}
\caption{The SEDs of a range of dusty galaxies with 
known redshifts (points), and two model spectra, that are 
chosen to accord with {\it IRAS} and {\it ISO} data (lines). There are 
four types of galaxies; low-redshift {\it IRAS} galaxies (I), sub-mm 
luminous galaxies (S), distant sources 
lensed by a 
foreground galaxy (L), and high-redshift 
radio-loud and radio-quiet AGN (H). The SED models include 
a dust emissivity index $\beta \simeq 1.5$ and a dust temperature 
$T_{\rm d} \simeq 40$\,K. Only the lensed galaxies and quasars, 
and some of the high-redshift AGN 
depart significantly 
from the template SEDs in the restframe mid-IR waveband.    
} 
\end{figure}

\section{The identification of optical counterparts} 

Unusually, because of the flat flux density--redshift relation illustrated 
in Fig.\,1, a measurement of the submm-wave flux density of a distant 
galaxy can be translated into a luminosity directly, independent of 
redshift, subject to an uncertain value of the dust temperature and 
emissivity index. However, a crucial step in understanding the 
nature of submm-selected galaxies is still the determination of their 
redshifts, individually and as a population. A problem arises because the 
SCUBA beam is 15\,arcsec in size at 850\,$\mu$m, and so there could 
easily be more than 5 plausible faint optical counterparts within 3\,arcsec 
of the centroid of a SCUBA detection in images as deep as the 
{\it Hubble Deep Field (HDF)} see Hughes et al.\ (1998) and Downes 
et al.\ (1999). Identifying the correct optical counterpart will in general be 
very difficult, especially as {\it a priori} submm-selected galaxies 
would be expected to have large internal extinctions and thus to be 
optically faint: see for example the source SMM\,J14009+0252 
(Ivison et al.\ 1999) and the brightest source in the SCUBA map of the 
{\it HDF} (Downes et al.\ 1999), which currently have no plausible 
optical counterparts. 

Sensitive radio observations are very useful for improving the 
chances of making a correct association, as the maximum resolution 
of the VLA at 1.4\,GHz is about 1\,arcsec. Any submm-selected galaxy 
should be detectable in a sufficiently deep radio image, based on the 
the observed low-redshift radio--far-IR correlation (Condon 1992), 
which links the synchrotron radio emission from supernova remnants 
and the submm-wave dust emission powered by young high-mass stars. 
The consequences for locating submm sources and estimating their 
redshifts are discussed by Carilli \& Yun (1999), Blain (1999b) and Carilli 
et al.\ (1999). This technique has been applied to a 
submm-selected sample by Smail et al.\ (1999b) and to a radio-selected, 
$K$-band faint sample by Barger, Cowie \& Richards (1999a). Note, 
however, that extremely deep radio observations are required in order to 
detect the current population of submm-selected 
sources. Only seven of Smail et al.'s fifteen SCUBA galaxies have 1.4-GHz 
detections, the rest only upper limits (Smail et al.\ 1999b). 

Plausible optical counterparts can be identified by overlaying the submm 
image with as many deep radio, near-IR and optical images as are available. 
Optical or near-IR spectroscopic redshifts are then required for the 
candidates. With such intrinsically faint sources, this 
is in general difficult, even using the largest telescopes (Ivison et al.\ 1998b; 
Barger et al.\ 1999c). The likelihood of an association being correct can be 
assessed based on photometric or positional arguments (Hughes et al.\ 1998; 
Smail et al.\ 1998); however, the final confirmation of the correct identification 
comes from the detection of molecular gas in interferometric mm-wave 
observations (Frayer et al.\ 1998, 1999). At 90-GHz the OVRO 
Millimeter Array has a resolution of several arcsec, a few times finer than 
SCUBA, and a bandwidth of 1\,GHz. Hence, if the redshift of a 
plausible counterpart is known to better than 1\%, ideally from near-IR 
spectroscopy of low-excitation lines, then the array can be tuned to the 
frequency of an appropriate CO rotation line at that redshift. If the line is 
detected at the position of the optical counterpart, then the presence of 
a large mass of molecular gas coincident in position and redshift ties 
together the optical and submm emission and confirms the identification. 

In only two cases has this procedure been completed, for 
SMM\,J02399$-$0136 (Ivison et al.\ 1998b; Frayer et al.\ 1998) at $z=2.8$ and 
SMM\,J14011+0252 (Ivison et al.\ 1999; Frayer et al.\ 1999) at $z=2.6$. Another 
source, SMM\,J02399$-$0134, has a very good candidate at $z=1.06$ 
(Barger et al.\ 1999c), with a blue ring galaxy morphology, distorted by 
the magnification of the cluster Abell 370, and a coincident 15-$\mu$m 
{\it ISO} detection (Soucail et al.\ 1999). 

With fluxes at or below the faintest detectable levels of radio and 
optical flux density, many of the counterparts of SCUBA sources 
necessarily remain enigmatic. The typical achromatic magnification 
by a factor of 2.5 experienced by background sources in the fields 
of rich clusters, and the associated expansion of the background sky 
by the same factor, has proved to be a great advantage for making 
follow-up observations. 

\begin{figure}[t]
\begin{center}
\plotfiddle{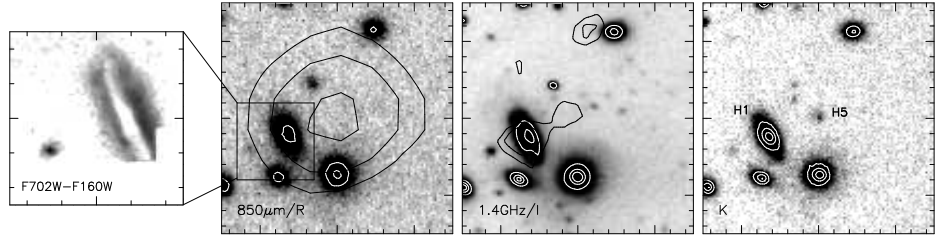}{2.1cm}{-90}{48}{48}{-205}{175}
\end{center} 
\caption{A multiwaveband view of the field of the submm galaxy 
SMM\,J09429+4658 (Smail et al.\ 1999a). The three larger images 
are 30\,arcsec on a side and are centred on the position of the 
centroid of the 850-$\mu$m detection. From left to right: An {\it HST} 
difference image showing the dust lane in the 
spiral galaxy H1; SCUBA 850-$\mu$m submm emission (black contours; 3, 6
and 9$\sigma$) overlaid on a Hale 5-m Gunn-$r$ image; VLA 1.4-GHz 
radio emission (black contours) overlaid on a deep Keck-II $I$-band 
image; and a UKIRT $K$-band image. The relative radio and optical astrometry
is accurate to less than 0.4\,arcsec.}   
\end{figure} 

\subsection{SMM\,J09429+4658: a case study} 

The SCUBA source SMM\,J09429+4658, for which an array of 
high-quality data is available, has proved to be a very interesting case 
study in making identifications of submm sources. In Fig.\,4 we show 
all the excellent multiwaveband data that has been compiled for this 
galaxy in the field of the $z=0.41$ cluster Cl\,0939+4713/A851. 
A partial two-colour {\it HST} image, a Hale $R$-band image, 
a very deep 1.4-GHz VLA map, a deep Keck-II I-band image and a UKIRT 
K-band image are available (Smail et al.\ 1999a). These images are 
extremely deep: after correcting for lens magnification, the 1$\sigma$ 
sensitivities are 3.6\,$\mu$Jy\,beam$^{-1}$ at 1.4-GHz and $I=27$. 

In the initial search for optical counterparts to the SCUBA galaxies 
(Smail et al.\ 1998), the dusty spiral galaxy H1 at $z=0.33$ in 
the foreground of the cluster was identified as a possible counterpart, 
despite being 7\,arcsec away from the SCUBA centroid and having 
no strong optical line emission (Barger et al.\ 1999c). 
When a deep 1.4-GHz VLA radio map was obtained, two radio sources 
were found within the SCUBA positional uncertainty: H1 (57\,$\mu$Jy) and 
a new source H5 (36\,$\mu$Jy), which has no optical counterpart 
but is aligned exactly 
with the SCUBA centroid. If H1 is the true counterpart, then unless 
its dust temperature is less than about 13\,K, it lies well below the 
far-IR--radio correlation at $z=0.33$; however, if H5 is the true counterpart, 
then its radio-submm flux ratio is typical of a 40-K dusty galaxy at 
$z \ge 3.5$ (Blain 1999b; Smail et al.\ 1999a). Unfortunately H5, lies off 
the edge of the {\it HST} image, and it has no optical counterpart in a 
very deep Keck-II $I$-band image. Only the acquisition 
of a deep UKIRT $K$-band image revealed a source at the 
H5 position. Keck NIRSPEC spectroscopy and, if a spectroscopic redshift 
is obtained, mm-wave observations of redshifted CO emission from 
H5 will hopefully confirm the identification over the 1999-2000 winter. 
The follow-up observations of this galaxy illustrate the importance of 
obtaining very deep multiwaveband data in order to identify 
SCUBA galaxies, especially radio and near-IR images.  

\section{The redshift distribution of submm-selected galaxies} 

The determination of the 
redshift distribution of submm-selected galaxies is crucial in order 
to constrain models of galaxy evolution (Blain et al.\ 1999c). There is 
considerable degeneracy between different models 
of the population of distant dusty galaxies that can all account for both the 
observed spectrum of mm, submm and far-IR background radiation and 
for the 450- and 850-$\mu$m counts of galaxies. The degeneracy 
can be broken most easily and definitively by determining the redshift 
distribution of the submm galaxies. Several different approaches have 
so far been taken to try and tackle this difficult problem. 

Smail et al.\ (1998) made plausible identifications of optical sources 
in deep {\it HST} and ground-based images of the fields of 
submm-selected galaxies behind seven rich clusters of galaxies. The 
lack of $V$-band dropout sources indicated that 80\% of the proposed 
counterparts were probably at $z<5$. Using Keck-II spectroscopy of this 
incomplete sample, Barger et al.\ (1999c) found a median redshift 
$\bar z \simeq 1.7$ for these counterparts. However, as illustrated by the 
detection of two EROs in the fields (Smail et al.\ 1999a), which were 
unknown at the time, this is likely to be a lower limit, as correct 
counterparts are more likely to be missing from the sample at higher 
redshifts. $\bar z \simeq 2.5$--3.0 is indicated by a Carilli \& Yun analysis 
of the radio--submm flux density ratios of the SCUBA galaxies 
derived from very deep VLA images of the fields 
(Smail et al.\ 1999b). 

Hughes et al.\ (1998) identified plausible faint optical counterparts to 
the brightest sources in the SCUBA {\it HDF} image, and deduced from 
photometric arguments that the redshifts of the sources were in the 
range $2 < z < 4$. Working from a deep 1.4-GHz VLA image, Richards 
(1999) suggested a systematic offset of the SCUBA astrometry, and a 
lower typical redshift. A subsequent high-resolution mm-continuum map 
of this source made using the IRAM interferometer (Downes et al.\ 1999) 
has provided a more accurate position, ruling out both of the counterparts 
suggested by Hughes and Richards, but suggesting that the source is 
gravitationally lensed by an elliptical galaxy. A very high redshift is 
potentially indicated for this source by the lack of a 
detectable optical 
counterpart in the {\it HDF} at this exact position. 

Using a variety of optical, mid-IR and radio data, Lilly et al.\ (1999) 
investigated the SCUBA sources detected in the CFRS fields 
(Eales et al.\ 1999), and found a significant fraction ($\simeq 30$\%) 
of plausible counterparts at $z < 1$. This is a lower redshift distribution 
than reported in other surveys. Misidentification of some of these $z<1$ 
galaxies remains a possibility; further follow-up observations should 
resolve this apparent discrepancy.  

Recently, Barger et al.\ (1999a) made SCUBA maps of the areas 
around the 1.4-GHz radio sources as faint as 40\,$\mu$Jy in the {\it HDF} 
flanking fields with no near-IR counterparts down to 
$K=20.5$. By using Carilli \& Yun's radio-submm redshift estimator 
they derived redshifts in the range $1 < z < 3$ for all six of the galaxies 
they detected using SCUBA. When the area of the submm maps and the 
number of detections are compared, it seems likely that there is a 
significant overlap between this population of near-IR-blank faint radio 
sources and the bright ($> 8$\,mJy) SCUBA galaxies. However, this 
radio--near-IR selection technique will tend to bias the sample as 
compared with a submm-selected sample, by enhancing the fraction 
of radio-loud objects and AGN.

At present, there are only two submm-selected galaxies with certain 
redshifts, SMM\,J02399$-$0136 and SMM\,J14011+0252 at redshifts 
$z=2.8$ and 2.6 respectively. This information alone suggests that the 
submm-selected population as a whole is probably at a significant 
redshift. Other techniques based on statistical arguments, tend to 
support the idea that $2 \le \bar z \le 4$ for the SCUBA galaxies, and 
that there is probably a tail of sources at higher redshifts. 

\section{The power source of submm-selected galaxies} 

It is clear from the observed SEDs of submm-selected galaxies that 
thermal emission from warm dust grains is being detected. However, 
it is an open question whether the power source heating the grains 
is the radiation from young stars or from the accretion disk of an 
AGN. The form of the broad-band submm and far-IR SED provides 
no information about this question. However, at shorter mid-IR wavelengths 
the slope of the continuum SED may tend to be harder in sources 
which are heated by AGN, for example compare the different slopes of 
Arp 220 and Mrk 231 in Fig.\,3. Also, the properties of PAH spectral 
features observed in nearby {\it IRAS} galaxies by the {\it ISO} satellite 
(Lutz et al.\ 1999), can provide an indication of the power source, because
the large molecules that produce them should not survive in 
the hard UV continuum radiation field expected in the core of an AGN. 
Lutz et al.\ (1999) conclude that 70\% of the far-IR radiation from these 
galaxies appears to be powered by star formation.

At present, direct observations of high-redshift submm-selected galaxies 
and of the X-ray background tend to support the continuation of this 
relative contribution out to high redshifts. One of the two best-studied 
SCUBA-selected galaxies (SMM\,J02399$-$0136; Ivison et al.\ 1998) shows 
clear signs of AGN activity, but the other (SMM\,J14011+0252; Ivison et al.\ 
1999) does not. Barger et al.\ (1999c) see evidence for high-excitation 
optical line emission, typical of AGN activity, in about 20\% of their 
relatively shallow Keck spectra of plausible optical counterparts to 
submm-selected galaxies (Smail et al.\ 1998). Note, however, that these 
are the easiest optical emission lines to detect, and so the AGN fraction 
in a complete sample may differ. A fraction of the optical counterparts targeted 
using the Keck will also certainly have been misidentified, leaving the 
true optical counterparts to the SCUBA galaxies unobserved.  
Almaini, Lawrence \& Boyle (1999) show that it is most plausible to 
explain the relative intensity of hard and soft X-ray background 
radiation by a population of high-redshift dust-enshrouded AGN, 
which can account for 10-20\% of the SCUBA sources.

It is not yet definite, but it seems likely that the majority of the emission 
making up the far-IR background and the counts of submm galaxies is 
powered by young high-mass stars, with a less significant 
contribution from AGN.

\section{Conclusions} 

\begin{enumerate} 
\item A significant population of high-redshift 
galaxies with powerful dust emission has been discovered since the 
commissioning of SCUBA, the first submm-wave bolometer array camera. 
\item The surface density of these galaxies is such that a significant 
fraction of the diffuse extragalactic background radiation detected by 
the {\it COBE} satellite has already been resolved into discrete sources. 
\item The SEDs of the SCUBA galaxies and other submm-luminous 
distant galaxies are generally consistent with a dust temperature 
$T_{\rm d} \simeq 40$\,K.
\item If there is a population of hotter submm-luminous sources, then 
these galaxies are less likely to be detected by SCUBA. At present 
there is little evidence that there is a significant population of such 
sources.
\item It is difficult and time consuming to identify and detect the current 
population of submm-selected galaxies in other wavebands. Only two 
such galaxies have so far been identified beyond reasonable doubt. 
\item It seems likely at present that star-formation activity dominates 
AGN accretion as a power source for submm-selected galaxies in the 
approximate proportion 2 or 3:1. The AGN-powered subset will hopefully 
soon be detected in high-resolution {\it Chandra} and {\it XMM} X-ray
observations. 
\end{enumerate}

\acknowledgements

The work described here is largely based on the results of the SCUBA Cluster 
Lens Survey, which has been lead by the authors, with collaborative 
support from Amy Barger, Jocelyn B\'ezecourt, Len Cowie, 
Aaron Evans, Dave Frayer, Allon Jameson, Jean-Francois Le Borgne, 
Malcolm Longair, Glenn Morrison, Frazer Owen, Nick Scoville, 
Paul van der Werf and Min Yun. AWB thanks the conference organizers 
for their hospitality and for support during the meeting, and Chris Carilli 
for helpful comments. This research has made use of the NASA/IPAC 
Extragalactic Database (NED) which is operated by JPL, Caltech, under 
contract with NASA.                                                          


%
%

%

\end{document}